\documentclass[journal]{IEEEtran}
\IEEEoverridecommandlockouts
\usepackage{cite}
\usepackage{amsmath,amssymb,amsfonts}
\usepackage{bm}
\usepackage{algorithmic,algorithm}
\usepackage{graphicx}
\usepackage{textcomp}
\usepackage{xcolor}
\usepackage{booktabs}
\usepackage{subcaption}
\usepackage{multirow}
\usepackage{makecell}
\usepackage{subfig}

\def\BibTeX{{\rm B\kern-.05em{\sc i\kern-.025em b}\kern-.08em
T\kern-.1667em\lower.7ex\hbox{E}\kern-.125emX}}
\begin{document}

\title{Spatial Computing Communications
for Multi-User Virtual Reality in Distributed Mobile Edge Computing Network
\author{Caolu Xu, Zhiyong Chen, \emph{Senior Member, IEEE}, Meixia Tao, \emph{Fellow, IEEE}, Li Song, \emph{Senior Member, IEEE}, \\and Wenjun Zhang, \emph{Fellow, IEEE}}

\thanks{The authors are with the Cooperative Medianet Innovation Center and the Department of Electronic Engineering, Shanghai Jiao Tong University, Shanghai 200240, China (e-mail: \{021034910015, zhiyongchen, mxtao, song\_li, zhangwenjun\}@sjtu.edu.cn). }}
\maketitle

\begin{abstract}
Immersive virtual reality (VR) applications impose stringent requirements on latency, energy efficiency, and computational resources, particularly in multi-user interactive scenarios. To address these challenges, we introduce the concept of spatial computing communications (SCC), a framework designed to meet the latency and energy demands of multi-user VR over distributed mobile edge computing (MEC) networks. SCC jointly represents the physical space, defined by users and base stations, and the virtual space, representing shared immersive environments, using a probabilistic model of user dynamics and resource requirements. The resource deployment task is then formulated as a multi-objective combinatorial optimization (MOCO) problem that simultaneously minimizes system latency and energy consumption across distributed MEC resources. To solve this problem, we propose MO-CMPO, a multi-objective consistency model with policy optimization that integrates supervised learning and reinforcement learning (RL) fine-tuning guided by preference weights. Leveraging a sparse graph neural network (GNN), MO-CMPO efficiently generates Pareto-optimal solutions. Simulations with real-world New Radio base station datasets demonstrate that MO-CMPO achieves superior hypervolume performance and significantly lower inference latency than baseline methods. Furthermore, the analysis reveals practical deployment patterns: latency-oriented solutions favor local MEC execution to reduce transmission delay, while energy-oriented solutions minimize redundant placements to save energy.
\end{abstract}

\section{Introduction}
Immersive communication has been identified by the International Telecommunication Union (ITU) as a key application scenario for sixth-generation (6G) mobile networks\cite{immersion_itu}. Among various modalities, virtual reality (VR) media is the most representative, exhibiting stringent requirements for high throughput, ultra-low latency, and intensive computational resources \cite{metaverse}. In addition to these, 6G networks are also expected to support real-time interactions and collaborative operations among multiple users, thereby driving the diversified development of future immersive applications.

Recently, various metaverse platforms and virtual collaboration tools have emerged, clearly demonstrating the growing demand for immersive and interactive communication. For instance, Roblox\cite{roblox} has been widely employed in the simulation of urban environment and the digital reconstruction of cultural heritage\cite{meta_games}. This metaverse game platform provides online gathering spaces to facilitate real-time user interaction. Furthermore, Roblox empowers users to construct personalized virtual worlds and make them accessible to others, reflecting rich interactivity and extensive user-generated content (UGC). Another example is Meta Horizon Workroom \cite{meta_workroom}, an immersive virtual workspace designed to enhance collaboration and interpersonal engagement among distributed team members. Utilizing devices like the Meta Quest headset\cite{meta_quest3} or conventional 2D screens, users can gather in a shared virtual environment to hold meetings, brainstorm ideas, and deliver presentations, thereby enhancing spatial co-presence and enabling real-time collaboration.

To gain a deeper understanding of real-time multi-user interactions in VR, this paper introduces spatial computing communications (SCC) for multi-user VR within a distribution mobile edge computing (MEC) network. Here, SCC primarily involve two spatial types: physical space and virtual space. Physical space refers to the real-world environment, capturing the relationship between users and their base stations (BS), while virtual space denotes the computer-generated VR environment, representing the interactive demands among multiple users. Users in different physical spaces can enter the same virtual space, while users in the same physical space can also enter different virtual spaces. The objective of this paper is to bridge these two spaces through resource deployment, aiming to enhance users' quality of experience (QoE).

\subsection{Related Work}
Recently, several works have explored how MEC contributes to VR. In \cite{Reliability_HetNets} and \cite{mmwave_compress}, mmWave and sub-6 GHz dual-connectivity is employed for wireless $360^{\circ}$ VR video transmission, jointly allocating edge and device resources. In \cite{two_timescale_VR}, a queue delay bound model is proposed for multiple VR video streams over a shared link, enabling long-term frame loss reduction and short-term adaptation to network fluctuations. Works such as \cite{proactive_caching_Tony_Quek} and \cite{cross_frame_prediction} leverage field of view (FoV) prediction to enable proactive video caching and partial computation offloading to MEC servers. Beyond the aforementioned works on $360^{\circ}$ VR video, recent studies address the interactive characteristics of VR in metaverse. Q-VR \cite{Q-VR} introduces a rendering solution through software-hardware co-design, motivated by the strong correlation between human visual perception and real-time hardware feedback. Task-oriented frameworks are proposed in \cite{JSAC_task_oriented_meta} and \cite{my_interactive_VR}, integrating sensing, communication, prediction, control, and rendering. Specifically, \cite{JSAC_task_oriented_meta} builds real-world robotic arms in the metaverse with accuracy and timeliness, while \cite{my_interactive_VR} presents an edge-device collaborative architecture with foreground-background separation to reduce device power consumption and mitigate user dizziness. 
 
Recent research has investigated the deployment of VR tasks in distributed MEC environments. In \cite{e2e_vr_social}, the end-to-end delay in mobile social VR networks is reduced through optimized bandwidth allocation in cellular systems. In \cite{QoE_vr_game_ADMM}, QoE-oriented edge assistance for multiple users of mobile VR games is formulated as a mixed integer quadratically constrained quadratic program (MIQCQP) problem and solved using the alternating direction method of multipliers (ADMM) \cite{Li_ADMM}. The study in \cite{INFOCOM_socialVR} investigates edge service entity placement (ESEP) for social VR by modeling a combinatorial optimization problem with multiple intertwined objectives, which is transformed into a graph cut problem solvable via maximum flow algorithms \cite{max_flow}. The authors in \cite{JSAC_dynamic_place} incorporate dynamic decision-making based on \cite{INFOCOM_socialVR} and jointly optimize operational cost and user performance, where model predictive control (MPC) \cite{MPC} is employed for online service placement within each prediction window. CO-TC \cite{CO_TC} proposes a collaborative offloading framework for VR clusters, using relation-aware graph embeddings empowered multi-agent reinforcement learning (MARL) \cite{MARL} to optimize both FoV task offloading and resource allocation. However, most existing studies focus on short-term decisions for ESEP. To avoid the overhead of frequent task migrations across the distributed MEC network, long-term optimization is crucial, particularly for caching decisions.

 Under resource constraints in physical space, deployment decisions form a discrete and non-convex combinatorial optimization (CO) problem. Prior studies have leveraged graph theory to handle unconstrained problems \cite{JSAC_gnn, INFOCOM_socialVR, JSAC_dynamic_place}, whereas variants of the max-flow algorithms suffer from significant computational overhead under complex constraints. Recently, artificial intelligence-generated content (AIGC) models such as GPT\cite{gpt4} and GLIDE\cite{GLIDE} have drawn considerable attention for their distribution-fitting and generative capabilities. In particular, diffusion models (DM) and their variants \cite{DDPM, DDIM, classfreeDM, CM} have demonstrated remarkable inference performance through iterative denoising. Motivated by these advances, recent research has explored DM based on graph neural networks (GNN) to address CO challenges. For instance, classical CO problems such as the traveling salesman problem (TSP) and maximal independent set (MIS) have been effectively solved using graph-based DMs in \cite{DIFUSCO, T2TCO, DIffUCO, Fast_T2T}. A review of DM for network optimization is provided in \cite{DM_reviewer}. 
 
 Building on these developments, we further consider a multi-objective combinatorial optimization (MOCO) problem targeting total system delay and energy consumption. This enables energy-efficient deployment under diverse latency requirements. In the domain of multi-objective optimization, reinforcement learning (RL)\cite{PPO} has demonstrated its ability to guide models toward superior Pareto fronts compared to traditional methods \cite{mo_ppo}. Moreover, RLs such as reinforcement learning from human feedback (RLHF) and group relative policy optimization (GRPO) have led to significant performance gains in large language models (LLM) \cite{GPT3, GRPO}, motivating researchers to explore the integration of RL with DMs. In \cite{DPOK}, authors derive a framework for applying RLHF to fine-tune text-to-image DMs. In \cite{mutiobj_co}, MOCO problem is converted into a single-objective problem via weighted scalarization, and a preliminary approach is developed to solve MOCO problems with RL guidance. However, these works have notable limitations. Firstly, since DMs require a large number of inference steps, their advantage over traditional solvers such as Gurobi \cite{Gurobi} becomes less pronounced in CO tasks. Secondly, the solution quality reported in \cite{mutiobj_co} is highly sensitive to the choice of objective weights. Therefore, we aim to design a preference specification mechanism as described in \cite{mo_mpo}, that supports adaptive normalization and remains invariant to scale.

\subsection{Contributions}
In this paper, we develop a spatial computing communication model and investigates resource deployment for multi-user VR in shared immersive environments over the distributed MEC network. The main contributions are summarized as follows:
\begin{itemize} 
    \item We design an SCC framework that bridge the physical and virtual spaces. The framework comprises two key components: (\romannumeral1) a probabilistic model that abstracts the dynamic VR users behaviors, including stochastic arrivals, departures, and transitions across virtual spaces; and (\romannumeral2) a multi-objective combinatorial optimization formulation with resource constraints, which jointly considers system delay and energy consumption to enable energy-efficient caching and computing under diverse latency requirements.
    \item We develop MO-CMPO, a multi-objective consistency model (CM) algorithm with policy optimization to solve the MOCO problem. The main innovations of the proposed MO-CMPO include:(\romannumeral1) reformulating coupled hard constraints into non-differentiable forms and solving them via a supervised learning task to reduce complexity; (\romannumeral2) designing a sparse graph neural network to capture the structural properties of the MOCO problem, together with an RL-based fine-tuning procedure for CM \cite{CM}; and (\romannumeral3) introducing preference weights to guide both supervised learning and RL fine-tuning, enabling adaptive normalization and scale invariance compared with conventional scalarization methods.
    \item We conduct extensive simulations using New Radio (NR) base station datasets from metropolitan area network (MAN) obtained through OpenCellid. Results demonstrate that the proposed algorithm outperforms conventional methods, while adapting effectively to user dynamics and varying multi-objective preferences under heterogeneous latency requirements.
\end{itemize}

The rest of this paper is organized as follows. Section \ref{System_Model} presents the system model of spatial computing communications for multi-user VR in distributed MEC network, and formulates the MOCO problem. Section \ref{mo_cmpo} reformulates primal problem, constructs a sparse GNN architecture, and proposes the MO-CMPO algorithm to obtain the solution. Simulation results are presented in Section \ref{simulation_results}. Section \ref{conclusion} concludes the paper.

\section{System Model and Problem Formulation}\label{System_Model}
\begin{figure}[t]
\centering{\includegraphics[width=0.5\textwidth]{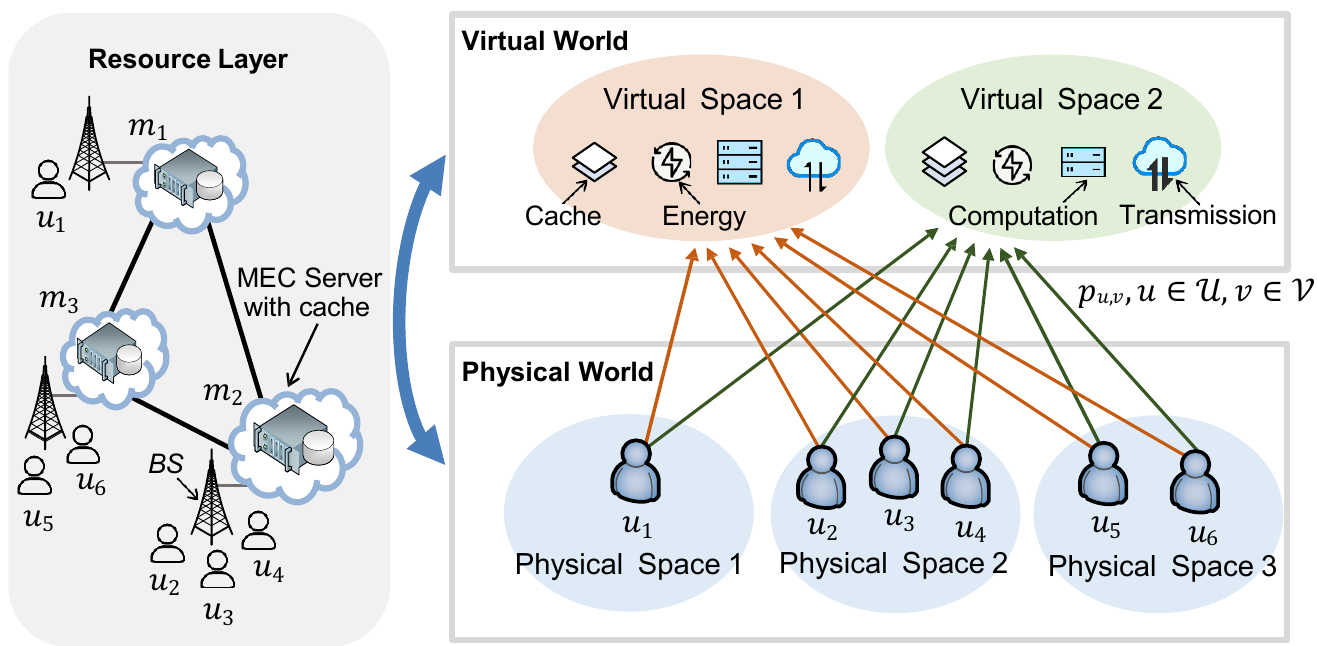}}
\caption{An illustration of the system model. }
\label{mapping}
\end{figure}

\renewcommand\arraystretch{1.3}
\begin{table}[h]
\caption{Key Notations}
\centering
\begin{tabular}{|>{\arraybackslash}m{1.6cm}|>{\raggedright\arraybackslash}m{5.7cm}|}
\noalign{\hrule height 0.4pt}
\textbf{Notation} & \textbf{Definition} \\
\noalign{\hrule height 0.4pt}
$\mathcal{U}$\,/\,$\mathcal{V}$\,/\,$\mathcal{M}$ & Set of users\,/\,virtual spaces\,/\,MECs \\
\noalign{\hrule height 0.4pt}
$c_v$\,/\,$E_v$\,/\,$h_v$\,/ $D_v$ & Level of cache size\,/\,basic energy consumption\,/ single-user computational workload\,/\,single-user data size for virtual space $v$\\
\noalign{\hrule height 0.4pt}
$p_{u,v}$ & Probability of user $u$ requesting service from virtual space $v$\\
\noalign{\hrule height 0.4pt}
$x_{u, v, m}$ & Computing placement decision for virtual space $v$ of user $u$ on MEC $m$\\
\noalign{\hrule height 0.4pt}
$y_{v, m}$ & Cache decision of virtual space $v$ on MEC $m$ \\
\noalign{\hrule height 0.4pt}
$C_m$\,/\,$F_m$\,/\,$N_m$& Cache size\,/\,computation frequency\,/\,maximum task number of MEC $m$\\
\noalign{\hrule height 0.4pt}
$d_{m,n}$\,/\,$e_{m,n}$ & Latency\,/\,energy consumption of transmitting synchronization information per virtual space from MEC $m$ to MEC $n$\\
\noalign{\hrule height 0.4pt}
$d_{u,m}$\,/\,$e_{u,m}$ & Latency\,/\,energy consumption of transmitting sensor information per virtual space from the local MEC of user $u$ to MEC $m$\\
\noalign{\hrule height 0.4pt}
$\kappa_{m,u}$\,/\,$\zeta_{m,u}$ & Latency\,/\,energy consumption per unit data volume for transmitting viewport frames from MEC $m$ to the local MEC of user $u$\\
\noalign{\hrule height 0.4pt}
$\kappa_{u}$\,/\,$\zeta_{u}$ & Latency\,/\,energy consumption per unit data volume for transmitting viewport frames from the local BS to the device of user $u$\\
\noalign{\hrule height 0.4pt}
$\tau^s$\,/\,$\tau^c$\,/\,$\tau^t$\,/\,$\tau^l$ & Latency of information synchronization\,/\,viewport computing\,/\,cross-MEC transmission\,/\,edge-device transmission in the system\\
\noalign{\hrule height 0.4pt}
$\varepsilon^m$\,/\,$\varepsilon^s$\,/\,$\varepsilon^c$\,/ $\varepsilon^t$\,/\,$\varepsilon^l$ & Energy consumption of virtual space maintenance\,/\,information synchronization\,/ viewport computing\,/\,cross-MEC transmission\,/ edge-device transmission in the system\\
\noalign{\hrule height 0.4pt}
\end{tabular}\label{notations}
\end{table}
As depicted in Fig. \ref{mapping}, the architecture supporting VR services consists of three components: the physical world, the virtual world and the resource layer. In the physical world, a BS with an MEC server can support several VR users. Users located under different BSs are considered to be in distinct physical spaces. Communication links exist between MEC servers, forming a distributed MEC network. In the virtual world, users in the same virtual space use the same VR service and interact with each other. The resource layer consists of computing, communication, and cache resources in the distributed MEC network. The connection between the physical world and the virtual world is established by mapping the virtual spaces of users to the distributed MEC resources. Key notations in this paper are summarized in Table \ref{notations}.

\subsection{Cost Factors}
The set of users, virtual spaces and MECs are denoted by $\mathcal{U}$, $\mathcal{V}$ and $\mathcal{M}$, respectively. We define levels for cached packet size, basic energy consumption, computational workload, and transmission data size. Then, virtual spaces are assigned to the corresponding levels. For each virtual space $v\in\mathcal{V}$, $c_v$ represents the cache size level, which corresponds to the stored data, e.g., vertices, textures, animations, illuminations, etc. 
The basic energy consumption level required to maintain virtual space $v$ is denoted by $E_v$. Since users within the same virtual space have similar load levels of computation and transmission, for virtual space $v\in\mathcal{V}$, the
floating-point operation (FLOP) level required to compute the viewport frame per user is indicated by $h_v$, and the data size level of viewport frame per user is denoted by $D_v$. The probability that user $u$ requests service from the virtual space $v$ is denoted by $p_{u,v}\in [0,1]$. Considering all virtual spaces of each user, the probability lies within the following range
\begin{align} \label{prob_vspace}
0\leq \sum_{v\in\mathcal{V}} p_{u,v}\leq1, \ \forall{u\in\mathcal{U}}.
 \end{align} $\sum_{v\in\mathcal{V}} p_{u,v}$ represents the probability that user $u$ exists in the VR system. Specifically, when $\sum_{v\in\mathcal{V}} p_{u,v}=0$, user $u$ leaves the interaction system, and when $\sum_{v\in\mathcal{V}} p_{u,v}=1$, user $u$ remains online. In contrast to previous works \cite{JSAC_gnn, INFOCOM_socialVR, JSAC_dynamic_place}, this probabilistic modeling provides a simple and flexible way to represent dynamic scenarios in which users switch between different virtual spaces, and enter or exit the VR system.

The computing placement decision for the virtual space $v$ of user $u$ on MEC $m\in\mathcal{M}$ is denoted by $x_{u,v,m}\in\{0,1\}$. $x_{u,v,m}=1$ indicates that the virtual space $v$ of user $u$ is deployed on MEC $m$, otherwise $x_{u,v,m}=0$. 
To ensure that the computing location is assigned to an MEC for user $u$ in virtual space $v$ when the entry probability is non-zero, the constraint is formulated as follows
\begin{align} \label{space constraint}
\sum_{m \in \mathcal{M}} x_{u,v,m} = \mathbb{I}^p_{u,v}, \  \forall{u\in\mathcal{U}}, \forall{v\in\mathcal{V}},
\end{align} 
where $\mathbb{I}^p_{u,v}=\mathbb{I}(p_{u,v}>0)$ indicates whether user $u$ has the probability of entering virtual space $v$. Specifically, if $\mathbb{I}^p_{u,v}=1$, i.e., $p_{u,v}>0$, one MEC in the physical space is allocated to compute the viewport of user $u$ in virtual space $v$. Conversely, if $\mathbb{I}^p_{u,v}=0$, no computing resource is allocated.
 
\begin{figure}[t]
    \hspace{0.02cm}
    \begin{minipage}[t]{0.29\linewidth}
        \centering
        \includegraphics[height=4.5cm, keepaspectratio]{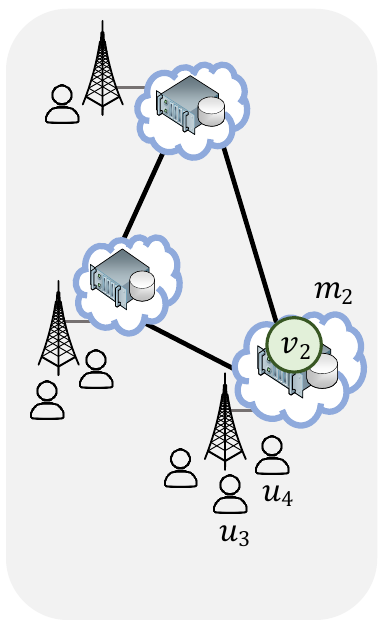}
        \subcaption{} \label{decision1}  
    \end{minipage}
    \hspace{0.05cm}
    \begin{minipage}[t]{0.29\linewidth}
        \centering
        \includegraphics[height=4.5cm, keepaspectratio]{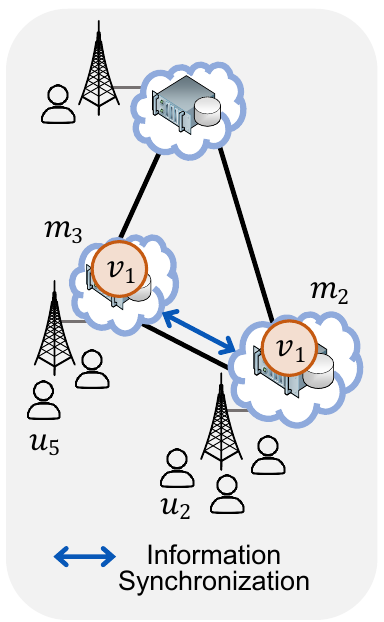}
        \subcaption{} \label{decision2} 
    \end{minipage}
    \hspace{0.05cm}
    \begin{minipage}[t]{0.29\linewidth}
        \centering
        \includegraphics[height=4.5cm, keepaspectratio]{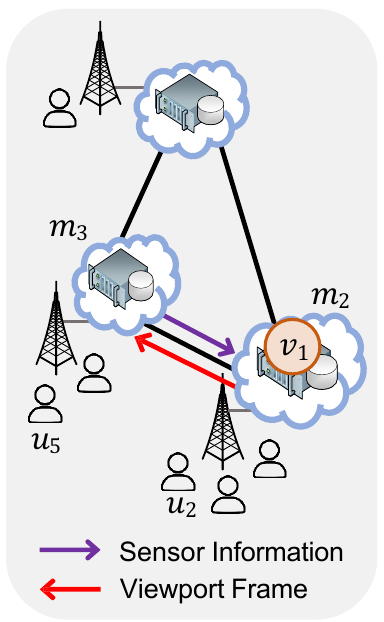}
        \subcaption{} \label{decision3}  
    \end{minipage}
    \caption{Mappings between virtual spaces and MEC nodes. (a) $u_3$ and $u_4$ are under the same BS with $m_2$. $p_{3,2}$ and $p_{4,2}$ are greater than zero. The corresponding decision variables are $y_{2,2}=1$, $x_{3,2,2}=1$, $x_{4,2,2}=1$. (b) $u_2$ and $u_5$ are connected to different BSs, with $m_2$ and $m_3$. $p_{2,1}$ and $p_{5,1}$ are greater than zero. The corresponding decision variables are $y_{1,2}=1$, $y_{1,3}=1$, $x_{2,1,2}=1$, $x_{5,1,3}=1$. (c) This scenario is the same as in (b). The corresponding decision variables are $y_{1,2}=1$, $y_{1,3}=0$, $x_{2,1,2}=1$, $x_{5,1,2}=1$.} 
    \label{decisions}
\end{figure}

We use $y_{v,m}\in\{0,1\}$ to denote whether MEC $m$ caches the virtual space $v$. $y_{v,m}=1$ indicates that the data packet of virtual space $v$ is cached on MEC $m$, otherwise $y_{v,m}=0$. There are certain constraints between computation and cache decisions. First, cache is a prerequisite for computation, and caching is not performed in the absence of computational demand. Additionally, as shown in Fig. \ref{decisions}(a), if multiple users are in the same virtual space $v$ on the same MEC $m$, the data packet of virtual space $v$ needs to be cached only once, i.e., $y_{v,m} = 1$ when $\sum_{u\in\mathcal{U}} x_{u,v,m} > 1$. Consequently, the relationship between computation and cache decisions can be formulated as
\begin{align} \label{cach_comp1}
y_{v,m}\geq x_{u,v,m}, \ \forall{u\in\mathcal{U}}, \forall{v\in\mathcal{V}}, \forall{m\in\mathcal{M}}, 
 \end{align} 
\begin{align} \label{cach_comp2}
y_{v,m}\leq \sum_{u\in\mathcal{U}}x_{u,v,m}, \ \forall{v\in\mathcal{V}}, \forall{m\in\mathcal{M}}. 
 \end{align}   

 \subsubsection{Packet Cache} 

The total cached data of the virtual spaces on MEC $m$ should not exceed its cache capacity $C_m$. The cache constraint can be expressed as
\begin{align} \label{constrant_cache}
\sum_{v\in\mathcal{V}}y_{v,m}c_v\leq C_m, \ \forall{m\in\mathcal{M}}. 
 \end{align}  
 
 \subsubsection{Virtual Space Maintenance}
For each MEC, caching virtual space $v$ incurs a basic maintenance energy consumption $E_v$. The total energy consumption for virtual space maintenance in the system can be represented by
 \begin{align} \label{varepsilon_m}
\varepsilon^m=\sum_{v\in\mathcal{V}}\sum_{m\in\mathcal{M}}y_{v,m} E_v.
 \end{align}

\subsubsection{Information Synchronization}
The mapping from virtual spaces to MEC nodes is one-to-many. As shown in Fig. \ref{decisions}(b), multiple users can access the same virtual space $v$ on more than one MEC node in the physical world, i.e., if $\sum_{u\in\mathcal{U}}\sum_{m\in\mathcal{M}}x_{u,v,m}>1$, $\sum_{m\in\mathcal{M}}y_{v,m}\geq1$, $\forall v\in\mathcal{V}$. When the same virtual space $v$ is deployed on different MECs, these MEC servers exchange synchronization information to update the state of all users within the virtual space. The total latency and total energy consumption of synchronization information transmission can be characterized by
\begin{align} \label{tau_s}
\tau^s=\sum_{v\in\mathcal{V}}\sum_{m\in\mathcal{M}}\sum_{n\in\mathcal{M}}d_{m,n}y_{v,m}y_{v,n},
 \end{align}
 \begin{align} \label{varepsilon_s}
\varepsilon^s=\sum_{v\in\mathcal{V}}\sum_{m\in\mathcal{M}}\sum_{n\in\mathcal{M}}e_{m,n}y_{v,m}y_{v,n},
 \end{align}
where $d_{m,n}$ and $e_{m,n}$ respectively represent the latency and energy consumption for transmitting synchronization information from MEC $m$ to MEC $n$ per virtual space\footnote{Compared to the large data size of viewport frame, the variation of the wired transmission cost in sensor information data size across different virtual spaces is negligible.}. Especially, when $m=n$, there is no synchronization cost and $d_{m,m}=0$ (or $e_{m,m}=0$). 

 \subsubsection{Viewport Computing} Based on the computing placement decision $x_{u,v,m}$, the viewport frame of user $u$ in virtual space $v$ is rendered on the corresponding MEC server. Considering the user request probability $p_{u,v}$ and the computational workload level of virtual spaces $h_v$, the total latency and total energy consumption for viewport frame computing in the system can be expressed as 
\begin{align} 
\tau^c=&\sum_{m\in\mathcal{M}}\sum_{u\in\mathcal{U}}\sum_{v\in\mathcal{V}}(p_{u,v}x_{u,v,m} h_v)/F_{m},\label{varepsilon_c}\\
\varepsilon^c=\xi&\sum_{m\in\mathcal{M}}\sum_{u\in\mathcal{U}}\sum_{v\in\mathcal{V}}(p_{u,v}x_{u,v,m}h_v)(F_{m})^2,\label{varepsilon_c}
 \end{align}
where $F_m$ denotes the computation frequency of MEC $m$, and $\xi$ represents the energy consumption coefficient related to hardware.

The maximum number of tasks that MEC $m$ can support is denoted as $N_m$. The number of computing tasks on each MEC should not exceed its maximum task capacity. The constraint on task number handled by MEC $m$ can be expressed as
\begin{align} 
\sum_{u\in\mathcal{U}}\sum_{v\in\mathcal{V}} x_{u,v,m} \leq N_m,\ \forall{m\mathrel{\mspace{-1.5mu}}\in\mathrel{\mspace{-2.2mu}}\mathcal{M}}. \label{N_total constraint}
 \end{align}
 
\subsubsection{Cross-MEC Transmission}
As depicted in Fig. \ref{decisions}(c), when the computation of virtual space $v$ is not deployed on the local MEC of user $u$, this incurs a cross-MEC transmission cost between the local MEC of user $u$ and the MEC where the virtual space computation is deployed. $d_{u,m}$ and $e_{u,m}$ respectively denote latency and energy consumption for transmitting sensor information from the local MEC of user $u$ to MEC $m$ per virtual space. $\kappa_{m,u}$ and $\zeta_{m,u}$ represent the latency and energy consumption per unit data volume for transmitting viewport frames from MEC $m$ to the local MEC of user $u$. In particular, when MEC $m$ is the local MEC to which user $u$ belongs, $d_{u,m}$, $e_{u,m}$, $\kappa_{m,u}$ and $\zeta_{m,u}$ are all equal to zero. Namely, the total latency and total energy consumption for cross-MEC transmission in the system are given by 
\begin{align}\label{tau_t}
\tau^t=& \sum_{m\in\mathcal{M}} \sum_{u\in\mathcal{U}} \sum_{v\in\mathcal{V}} p_{u, v} x_{u, v, m} (d_{u,m}+\kappa_{m,u} D_v),
\end{align}
\begin{align}\label{varepsilon_t}
\varepsilon^t=& \sum_{m\in\mathcal{M}} \sum_{u\in\mathcal{U}} \sum_{v\in\mathcal{V}} p_{u, v} x_{u, v, m} (e_{u,m}+\zeta_{m,u} D_v).
\end{align}

\subsubsection{Edge-Device Transmission}
Wireless transmission occurs between user devices and their local BS. Since the data volume of sensor information is much smaller than that of viewport frame, the wireless transmission cost of sensor information can be neglected. 
The latency and energy consumption per unit data volume for transmitting viewport frames from the local BS to the device of user $u$ are respectively denoted as $\kappa_{u}$ and $\zeta_{u}$. The total latency and total energy consumption of edge-device transmission can be expressed as

\begin{align}
\tau^l= &\sum_{u\in\mathcal{U}}\sum_{v\in\mathcal{V}}  p_{u, v} \kappa_{u} D_v,\label{tau_l}\\
\varepsilon^l=&\sum_{u\in\mathcal{U}} \sum_{v\in\mathcal{V}} p_{u, v}\zeta_{u} D_v.\label{varepsilon_l}
\end{align}

$\boldsymbol{x}\triangleq (x_{u,v,m})_{u\in\mathcal{U}, v\in\mathcal{V}, m\in\mathcal{M}}$ and $\boldsymbol{y}\triangleq (y_{v,m})_{v\in\mathcal{V}, m\in\mathcal{M}}$ are binary optimization variables.
The edge-device transmission costs (\ref{tau_l}) and (\ref{varepsilon_l}) are independent of $\boldsymbol{x}$ and $\boldsymbol{y}$. $\tau^l$ and $\varepsilon^l$ can thus be excluded from the latency and energy consumption objectives. Consequently, the total latency and total energy consumption in the system are given by
\begin{align}
T(\boldsymbol{x}, \boldsymbol{y}) = &\tau^s+\tau^c+\tau^t,\label{obj_T}\\
E(\boldsymbol{x}, \boldsymbol{y}) = \varepsilon^m&+\varepsilon^s+\varepsilon^c+\varepsilon^t. \label{obj_E}
\end{align}

\subsection{Optimization Problem}
In multi-user VR interactive scenarios, reducing latency is a key metric for ensuring quality of service. At the same time, improving energy efficiency of the MEC network is also essential. Accordingly, we formulate the optimization problem to minimize the latency and energy consumption simultaneously as following
\begin{align}
\mathcal{P}1:\ 
\min_{     
        \boldsymbol{x},\boldsymbol{y}}& \ [T(\boldsymbol{x},\boldsymbol{y}),E(\boldsymbol{x},\boldsymbol{y})] \nonumber\\
\text { s.t. }\ &(\ref{space constraint}), (\ref{cach_comp1}), (\ref{cach_comp2}), (\ref{constrant_cache}), (\ref{N_total constraint}).\nonumber
\end{align}\label{p1}
This problem is a multi-objective combinatorial optimization problem. A user-local MEC deployment strategy can reduce latency but leads to replication and synchronization, introducing increased energy consumption. Reducing the number of synchronized virtual spaces can lower total energy consumption, but it increases cross-MEC transmission and results in longer latency. Thus, we aim to find Pareto optimal solutions between the two conflicting objectives. In the following sections, we present how to obtain a set of Pareto optimal decisions for problem $\mathcal{P}1$.

\section{Multi-Objective Consistency Model based Policy Optimization Algorithm}\label{mo_cmpo}
Specifically, constraints (\ref{cach_comp1}) and (\ref{cach_comp2}) in $\mathcal{P}1$ are hard constraints, which tightly couple the variables $\boldsymbol{x}$ and $\boldsymbol{y}$. Directly embedding them into neural networks would significantly increase model complexity. To address this, we first reformulate $\mathcal{P}1$ by eliminating $\boldsymbol{y}$ through the following lemma.

\textbf{Lemma 1.} \textit{For all \(v \in \mathcal{V}\), \(m \in \mathcal{M}\), define} 
\begin{align} \label{I_x}
\mathbb{I}^x_{v,m} = &\mathbb{I}\Big(\sum_{u\in\mathcal{U}} x_{u,v,m} > 0\Big), \textit{ then}\\
\mathbb{I}^x_{v,m}\textit{ is a valid substi}&\textit{tution for }y_{v,m}\textit{, such }(\ref{cach_comp1})\textit{ and }(\ref{cach_comp2})\textit{ hold}.\nonumber
\end{align}

\textit{Proof.} We show that the indicator function $\mathbb{I}^x_{v,m}$ satisfies constraints (\ref{cach_comp1}) and (\ref{cach_comp2}). 

Step 1. Show that $\mathbb{I}^x_{v,m}$ satisfies $\mathbb{I}^x_{v,m}\geq x_{u,v,m}, \forall u \in \mathcal{U}$: If $x_{u,v,m} = 1$, then $\sum_{u \in \mathcal{U}} x_{u,v,m} > 0$, implying $\mathbb{I}^x_{v,m} = 1$. Hence, the inequality holds.

Step 2. Show that $\mathbb{I}^x_{v,m} \leq \sum_{u \in \mathcal{U}} x_{u,v,m}$: If $\sum_{u \in \mathcal{U}} x_{u,v,m} = 0$, then $\mathbb{I}^x_{v,m} = 0$, satisfying the inequality.
If $\sum_{u \in \mathcal{U}} x_{u,v,m} > 0$, then $\mathbb{I}^x_{v,m} = 1 \leq \sum_{u \in \mathcal{U}} x_{u,v,m}$, so the constraint still holds.   

In conclusion, $\mathbb{I}^x_{v,m}$ is equivalent to $y_{v,m}$. The optimization problem preserves the equivalence after reformulation.\hfill $\square$

Compared to satisfying hard coupling constraints, neural networks better suited to avoiding non-differentiable optimization problems through flexible network design. Therefore, we utilize Lemma 1 to eliminate the auxiliary optimization variable $\boldsymbol{y}$ and its associated constraints (\ref{cach_comp1})-(\ref{cach_comp2}) by substituting $y_{v,m}$ with $\mathbb{I}^x_{v,m}$. This leads to the following reformulated problem,
\begin{align}
\mathcal{P}2:\ 
\min_{     
        \boldsymbol{x}}& \ [T(\boldsymbol{x}),E(\boldsymbol{x})] \nonumber\\
\text { s.t. }\ &(\ref{space constraint}), (\ref{N_total constraint}),\nonumber\\
&\sum_{v\in\mathcal{V}}\mathbb{I}^x_{v,m}c_v\leq C_m, \ \forall{m\in\mathcal{M}}.
\end{align}\label{p2}
To address this MOCO problem, we propose the MO-CMPO algorithm as an alternative to traditional weighted scalarization methods. The proposed algorithm leverages preference weights to guide the supervised learning of CM and the RL fine-tuning.

\subsection{Supervised Learning with Consistency Model}
\begin{figure*}[t]
\centering{\includegraphics[width=0.7\textwidth]{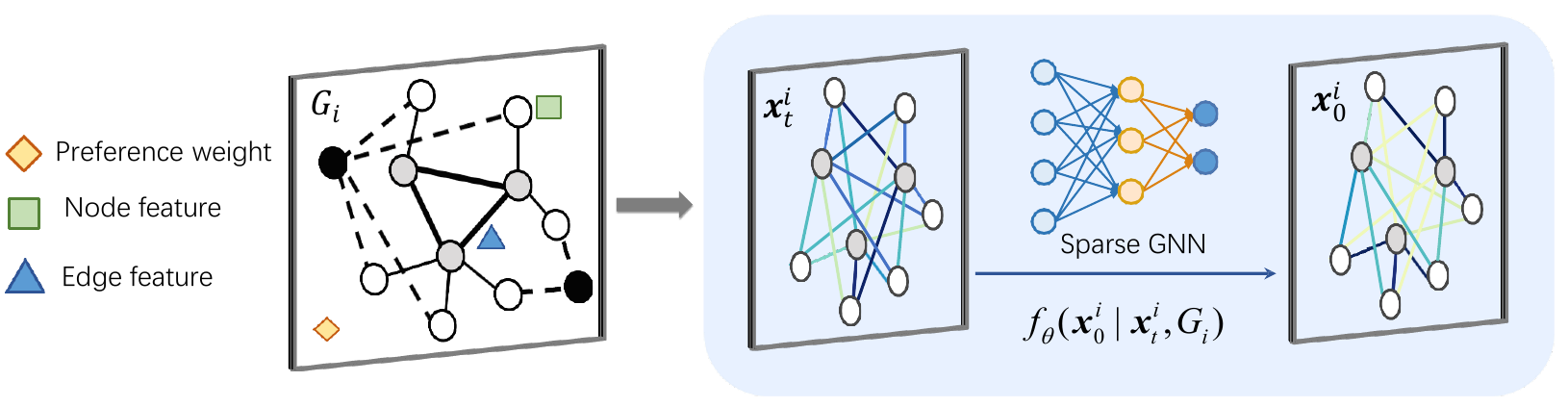}}
\caption{An illustration of proposed multi-objective optimization with consistency model.}
\label{cm}
\end{figure*}

DMs have demonstrated strong distribution modeling capabilities and achieved remarkable success in image generation\cite{DDPM} and discrete optimization\cite{DM_reviewer}. Inspired by these advances, we adopt a state-of-the-art consistency model as the base model for solving $\mathcal{P}$2. Compared with standard DMs, CM enables faster inference with significantly fewer steps, effectively mitigating the long inference latency caused by the large number of denoising iterations. An illustration of multi-objective optimization using the CM architecture in Fig. \ref{cm}.

At the training stage, the problem $\mathcal{P}$2 is formulated as a sparse GNN, denoted $G$, which consists of nodes and edges. The nodes in $G$ comprise three types: user set $\mathcal{U}$, MEC set $\mathcal{M}$, and virtual space set $\mathcal{V}$. The user node features include $\kappa_u$ and $\zeta_u$; the MEC node features include $C_m, F_m, $ and $N_m$; and the virtual space node features include $c_v, E_v, h_v,$ and $D_v$. The edge features comprise: the probability $p_{u,v}$ that user $u$ requests virtual space $v$; the MEC to which the user physically belongs; the inter-MEC latency $d_{m,n}$ and the energy costs $e_{m,n}$ for transmitting synchronization information; the coefficients related to latency and energy between users and MECs $d_{u,m}, e_{u,m}, \kappa_{m,u}, \zeta_{m,u}$.

For the multi-objective problem $\mathcal{P}2$, we generalize the two-objective problem to the case with $I$ objectives. A preference weight vector is denoted as $\boldsymbol{w}$, 
\begin{align} \label{preference weight}
\boldsymbol{w}\mathrel{\mspace{-3mu}}= \mathrel{\mspace{-3mu}}[w_1, w_2, \cdots, w_I] \in \mathbb{R}^I\mathrel{\mspace{-3mu}},\text{with } \sum_{i=1}^{I} w_i= 1,w_i\geq 0.
\end{align}
The single-objective optimization in $\mathcal{P}1$ is an integer quadratic programming (IQP), which can be directly solved using Gurobi\cite{Gurobi}. The solutions obtained from Gurobi are used as labels for supervised learning of the CM. Each single-objective solution $\boldsymbol{x}_0^i$ corresponds to $\boldsymbol{w}=\boldsymbol{e}_i$, the one-hot vector with 1 at the $i$-th position. Each $\boldsymbol{x}_0^i$ with a corresponding $\boldsymbol{w}=\boldsymbol{e}_i$ is embedded into the global graph $G$, resulting in an augmented graph representation $G_i$, $i\in\{1,2,\cdots,I\}$. A consistency model $\theta$ is trained to generate solutions $\boldsymbol{x}_0^i$ conditioned on the latent variable $\boldsymbol{x}_t^i$ and the graph $G_i$, i.e., $f_{\theta}(\boldsymbol{x}_0^i \mid \boldsymbol{x}_t^i, G_i)$. 

In the specific case of $\mathcal{P}2$, where $I=2$, the solutions to the objectives $T(\boldsymbol{x},G)$ and $E(\boldsymbol{x},G)$ are denoted by $\boldsymbol{x}_0^1$ and $\boldsymbol{x}_0^2$, respectively. Single-objective solutions are obtained by solving optimizations in $\mathcal{P}1$ using Gurobi, and correspond to the preference weights $\boldsymbol{w} = [1, 0]$ and $[0, 1]$. 

\subsubsection{Noising Process} The noising process follows the discrete diffusion modeling framework. Starting from the initial solution $\boldsymbol{x}_0^i$, noise is gradually introduced to generate a sequence of latent variables $\boldsymbol{x}_{1:T}^i = \boldsymbol{x}_1^i, \boldsymbol{x}_2^i, \cdots, \boldsymbol{x}_T^i$.  The noising process can be expressed as $q(\boldsymbol{x}_{1:T}^i \mid \boldsymbol{x}_0^i) = \prod_{t=1}^T q(\boldsymbol{x}_t^i \mid \boldsymbol{x}_{t-1}^i)$. At step $t$, the latent variable $\boldsymbol{x}_t^i$ is obtained by multiplying the previous state $\boldsymbol{x}_{t-1}^i$ with a forward transition probability matrix $\boldsymbol{Q}_t=\left[\begin{array}{cc}\beta_t & 1-\beta_t \\ 1-\beta_t & \beta_t\end{array}\right]$, and $\beta_t \in [0,1]$ \cite{discrete_DM}. Since all entries of $\mathbf{Q}_t$ are strictly positive, the stationary distribution converges to the uniform distribution. The noising process and marginal probability at step $t$ is defined as
\begin{align} \label{noising process}
q\left(\boldsymbol{x}_t^i \mid \boldsymbol{x}_{t-1}^i\right)=\operatorname{Cat}\left(\boldsymbol{x}_t^i ; \boldsymbol{p}=\boldsymbol{x}_{t-1}^i \boldsymbol{Q}_t\right),\end{align}
\begin{align} \label{marginal probability}
q\left(\boldsymbol{x}_t^i \mid \boldsymbol{x}_{0}^i\right)=\operatorname{Cat}\left(\boldsymbol{x}_t^i ; \boldsymbol{p}=\boldsymbol{x}_{0}^i \overline{\boldsymbol{Q}}_t\right),\end{align}
where $\operatorname{Cat}(\boldsymbol{x}^i; \boldsymbol{p}), i\in\{1,2\}$ is a categorical distribution over variables, and $\overline{\boldsymbol{Q}}_t=\boldsymbol{Q}_1 \boldsymbol{Q}_2 \cdots \boldsymbol{Q}_t$.

\subsubsection{Denoising Process} For a given $G_i$, every point along the trajectory yields the same optimal solution $\boldsymbol{x}_0^i$, i.e., $
f_\theta\left(\boldsymbol{x}_0^i \mid\boldsymbol{x}_{t}^i, G_i\right) = f_\theta\left(\boldsymbol{x}_0^i \mid\boldsymbol{x}_{t'}^i, G_i\right) = \delta\left(\boldsymbol{x} - \boldsymbol{x}_0^i\right)$\cite{CM}, where $t$ and $t'$ are distinct steps from different trajectories, and $\delta(\cdot)$ is the unit impulse distribution. Denote $H(\cdot)$ as binary cross entropy, then the loss function of consistency model is $\mathcal{L}_0(\theta) = \mathbb{E}[\sum_{i\in\{1,2\}}H(f_\theta\left(\boldsymbol{x}_0^i \mid\boldsymbol{x}_{t}^i, G_i\right), f_\theta\left(\boldsymbol{x}_0^i \mid\boldsymbol{x}_{t'}^i, G_i\right))]$. Based on the triangle inequality of distance measures, the upper bound of $\mathcal{L}_0(\theta)$ is $\mathcal{L}_{CM}(\theta)$, which is denoted by
\begin{align} \label{CM_loss}
\mathcal{L}_{CM}(\theta) = \mathbb{E}\Big[\sum_{i\in\{1,2\}}H\left(f_\theta\left(\boldsymbol{x}_0^i \mid \boldsymbol{x}_{t}^i, G_i\right), \delta\left(\boldsymbol{x} - \boldsymbol{x}_0^i\right)\right)\nonumber\\+ H\left(f_\theta\left(\boldsymbol{x}_0^i \mid \boldsymbol{x}_{t'}^i, G_i\right), \delta\left(\boldsymbol{x} - \boldsymbol{x}_0^i\right)\right)\Big],\end{align}
where $\boldsymbol{x}_{t}^i \sim \operatorname{Cat}\left(\boldsymbol{x}_{t}^i; \boldsymbol{p}=\boldsymbol{x}_0^i\overline{\boldsymbol{Q}}_{t}\right)$. The denoising time steps are chosen as a subset of $\{1,2,\cdots,T\}$ according to the cosine scheduler, defined as $t=\left\lfloor\cos \left(\frac{1-n\pi}{2}\right) \cdot T\right\rfloor$ in DDIM\cite{DDIM}. The training stage is summarized in Algorithm \ref{CM_training_alg}. 

At the test stage, $K$ step consistency sampling is presented in Algorithm \ref{CM_sampling_alg}.

\begin{algorithm}[t]
\hspace*{0.08in}{\bf \small{Input:}} \small{Initial consistency model network $\theta$, graph set $\mathcal{G}$, supervised training solution set $\Phi$, hyper-parameter $T$ and $\alpha$;}\\
\hspace*{0.04in} {\bf \small{Output:}} \small{Trained consistency model $f_{\theta}(\cdot)$;}
\caption{Consistency Training}
\label{CM_training_alg}
\begin{algorithmic}[1]
\WHILE{the training stop is not met}
\STATE Randomly sample $t$ from $Uniform(\{1,...,T\})$, $t'=\alpha T$;
\STATE Randomly sample $G_i$ and $\boldsymbol{x}_i$ from $\mathcal{G}$ and $\Phi$;
\STATE Sample $\boldsymbol{x}_{t}^i\sim \operatorname{Cat}\left(\boldsymbol{x}_{t}^i; \boldsymbol{p}=\boldsymbol{x}_0^i\overline{\boldsymbol{Q}}_{t}\right)$, \\\quad \ \ and $\boldsymbol{x}_{t'}^i\sim \operatorname{Cat}\left(\boldsymbol{x}_{t'}^i; \boldsymbol{p}=\boldsymbol{x}_0^i\overline{\boldsymbol{Q}}_{t'}\right)$ according to (\ref{marginal probability});
\STATE Take gradient descent step $\nabla_\theta \left(L_{CM}(\theta)\right)$ according to (\ref{CM_loss});
\ENDWHILE
\end{algorithmic}
\end{algorithm}

\begin{algorithm}[t]
\hspace*{0.08in}{\bf \small{Input:}} \small{Trained consistency model $f_{\theta}(\cdot)$, graph $G_i, i\in\{1,2\}$, time steps $t_1>t_2>\cdots> t_{K-1}$;}\\
\hspace*{0.04in} {\bf \small{Output:}} \small{Solution $\boldsymbol{x}_0$;}
\caption{Consistency Sampling}
\label{CM_sampling_alg}
\begin{algorithmic}[1]
\STATE Randomly sample $\boldsymbol{x}_T$ from the uniform distribution;
\STATE $\boldsymbol{x}_0^{i,0}\sim f_{\theta}(\boldsymbol{x}_0^{i,0} \mid \boldsymbol{x}_T, G_i)$;
\FOR{${k} = 1, 2, \dots, K-1$}
    \STATE Sample $\boldsymbol{x}_{t_k}^{i}\sim \operatorname{Cat}\left(\boldsymbol{x}_{t_k}^i; \boldsymbol{p}=\boldsymbol{x}_0^{i,k}\overline{\boldsymbol{Q}}_{t_k}\right)$;
    \STATE $\boldsymbol{x}_0^{i,k+1}\sim f_{\theta}(\boldsymbol{x}_0^{i,k} \mid \boldsymbol{x}_{t_k}, G_i)$;
\ENDFOR
\end{algorithmic}
\end{algorithm}

\subsection{Fine-tuning with Reinforcement Learning}
\begin{figure*}[t]
\centering{\includegraphics[width=0.7\textwidth]{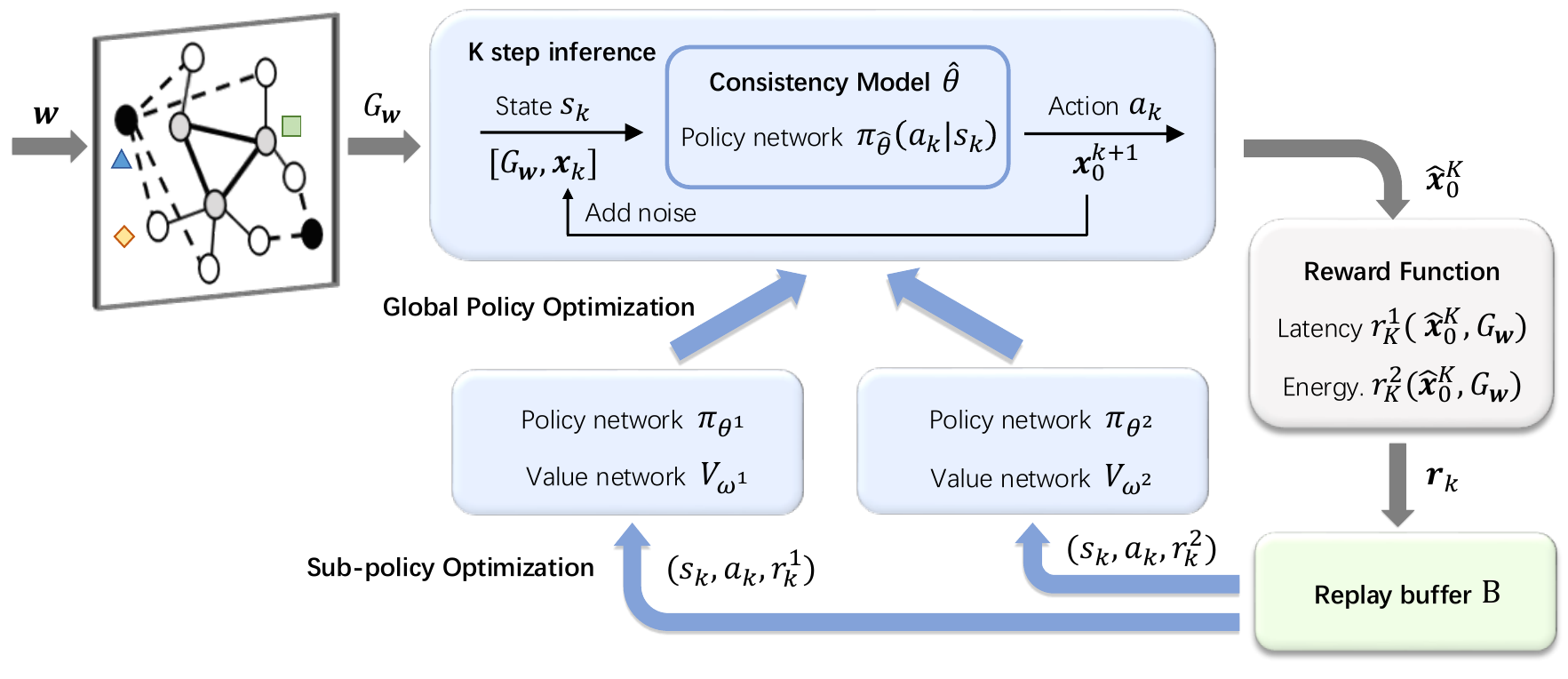}}
\caption{The structure of proposed multi-objective RL fine-tuning algorithm.}
\label{rl}
\end{figure*}

RL enables efficient and robust decisions in dynamic environment. Based on the pre-trained consistency model, we design a multi-objective RL algorithm to obtain the Pareto front of the MOCO problem. An illustration of multi-objective RL fine-tuning algorithm is provided in Fig. \ref{rl}. 

 The Markov decision process (MDP) is defined as $(\mathcal{S}, \mathcal{A}, \mathcal{R})$, where $\mathcal{S}$ denotes the state space, $\mathcal{A}$ denotes the action space, and $\mathcal{R}$ denotes the reward space. For notational simplicity, the index $i$ of the single objective is omitted and $k$ is used in place of $t_k$ in Algorithm~\ref{CM_sampling_alg} throughout subsequent sections. Let $k\in\{0,1,\dots,K{-}1\}$. Specifically, the initialized $\boldsymbol{x}_T$ from a uniform distribution corresponds to $k=0$. For the $k$-th inference step of consistency model, the state is given by $s_k = [\boldsymbol{x}_k, G] \in \mathcal{S}$, the action $a_k\in \mathcal{A}$ is $\boldsymbol{x}_{0}^{k+1}$, and the reward $\boldsymbol{r}_{k+1} \in \mathcal{R}$ is denoted by
 \begin{align} \label{rewards}
\boldsymbol{r}_{k+1} = \begin{cases}[-T(\hat{\boldsymbol{x}}_0^{K}, G), -E(\hat{\boldsymbol{x}}_0^{K}, G)], & \text{if } k=K{-}1, \\
[0, 0], & \text{otherwise},
\end{cases}
\end{align}
where $\hat{\boldsymbol{x}}_0^K$ is a feasible neighbor of the inference output $\boldsymbol{x}_0^K$, obtained via a greedy algorithm to maximize the sum of output probability and enforce all constraints in the combinatorial optimization problem $G$. Since consistency sampling does not guarantee feasibility, $\boldsymbol{x}_0^K$ is projected onto the constraint-satisfying space to yield $\hat{\boldsymbol{x}}_0^K$.

The RL fine-tuning of the CM is performed iteratively. $\hat{\theta}_j$ denotes CM policy parameters of the $j$-th iteration. The initial policy parameters $\hat{\theta}_0$ are set to those of the pre-trained multiobjective CM $\theta$. The Pareto solution set $\Lambda_0$ is initialized as empty. The following presents a detailed description of each RL fine-tuning iteration.

\subsubsection{Environment Interaction and Buffer Data Sampling}
We model the preference vector $\boldsymbol{w}$ using a Dirichlet distribution, which naturally satisfies the constraints in (\ref{preference weight}). Its probability density function (PDF) is given by,
 \begin{align} \label{Dirichlet}
\boldsymbol{w}\sim\frac{1}{\operatorname{Beta}(\boldsymbol{\chi})} \prod_{i=1}^I w_i^{\chi_i-1},
\end{align}
where the multivariate Beta function $\operatorname{Beta}(\boldsymbol{\gamma})$ is given by
 \begin{align} \label{beta}
\operatorname{Beta}(\boldsymbol{\chi})=\frac{\prod_{i=1}^I \Gamma\left(\chi_i\right)}{\Gamma\left(\sum_{i=1}^I \chi_i\right)},
\end{align}
and $\Gamma(\cdot)$ denotes the Gamma function. When each $\chi_i$ is a positive integer, $\Gamma(\chi_i) = (\chi_i{-}1)!$. 

To enable unbiased and uniform exploration across all objective directions, we set $\boldsymbol{\chi} = [1, \dots, 1]$, such that the preference weights $\boldsymbol{w}$ are sampled from a uniform Dirichlet distribution over the $(I-1)$-dimensional probability simplex. The normalization constant simplifies to $\operatorname{Beta}(\boldsymbol{1}) = 1/{(I{-}1)!}$.
In this paper, we focus on the bi-objective case ($I = 2$). The Dirichlet distribution (\ref{Dirichlet}) reduces to the Beta distribution $\operatorname{Beta}(1, 1)$, hence $\boldsymbol{w}$ corresponds to a uniform distribution 
 \begin{align} \label{weight_buffer}
\boldsymbol{w} = [w, 1 - w], \ \ \text{where}\ w\sim\operatorname{Uniform}(0, 1).
\end{align}

An instance $G$ is drawn from the combinatorial optimization problem set, and $\boldsymbol{w}$ is embedded in the global graph $G_{\boldsymbol{w}}$. Based on the current policy $\pi_{\hat{\theta}_j}$, we perform $K$ step inference in Algorithm \ref{CM_sampling_alg}. $\boldsymbol{x}_k$ is iteratively sampled from
\begin{align}\label{samplex_buffer}
\boldsymbol{x}_{k}\mathrel{\mspace{-3mu}}\sim\mathrel{\mspace{-3mu}} \operatorname{Cat}\left(\boldsymbol{x}_{k}; \boldsymbol{p}\mathrel{\mspace{-3mu}} = \mathrel{\mspace{-3mu}}\boldsymbol{x}_0^k \overline{\boldsymbol{Q}}_{k}\right)\mathrel{\mspace{-5mu}}, \mathrel{\mspace{-3mu}}\boldsymbol{x}_0^{k+1}\mathrel{\mspace{-3mu}}\sim\mathrel{\mspace{-3mu}} f_{\hat{\theta}_j}(\boldsymbol{x}_0^{k+1}\mathrel{\mspace{-3mu}}\mid \mathrel{\mspace{-3mu}}\boldsymbol{x}_{k}, G_{\boldsymbol{w}}),
\end{align}
where $k \in \{0, 1, \dots, K{-}1\}$, and update the state $\boldsymbol{s}_k = (\boldsymbol{x}_k, G_{\boldsymbol{w}})$, the action $a_k=\boldsymbol{x}_{k+1}$ at each step. After the inference, we apply a greedy algorithm to obtain a feasible solution $\hat{\boldsymbol{x}}_0^K$. According to (\ref{rewards}), we obtain the rewards on the two single-objective functions as $r^1_{K} = -T(\hat{\boldsymbol{x}}_0^K,G_{\boldsymbol{w}})$ and $r^2_{K} = -E(\hat{\boldsymbol{x}}_0^K,G_{\boldsymbol{w}})$, and then update the corresponding reward vector $\boldsymbol{r}_{K}$. $(s_k, a_k, \boldsymbol{r}_{k+1})$ is stored in the replay buffer $B_j$. This sampling process is repeated until $B_j$ is filled.

\subsubsection{Separate Policy Optimization for Individual Objectives}
An actor-critic architecture based on proximal policy optimization (PPO) is adopted to optimize the policy along each individual objective direction. $\theta^i_j$ and $\omega^i_j$, $\forall i\in\{1,2\}$, respectively denote the policy parameters and cirtic parameters of the $i$-th objective function in the $j$-th iteration. Parameters for each objective are updated by minimize the following function,
\begin{align}\label{PPO}
    \mathcal{L}(\theta^i_j, \omega^i_j)=\mathcal{L}_{\text {policy}}(\theta^i_j)-\eta_1 \mathcal{H}(\theta^i_j)+\eta_2 \mathcal{L}_{\text {value}}(\omega^i_j),
\end{align}
where $\mathcal{L}(\theta^i_j, \omega^i_j)$ consists of three components: the cumulative reward $\mathcal{L}_{\text{policy}}(\theta^i_j)$, the entropy of the action distribution $\mathcal{H}(\theta^i_j)$, and the prediction error of the critic network $\mathcal{L}_{\text{value}}(\omega^i_j)$. $\eta_1$ and $\eta_2$ are positive constants. $\mathcal{L}_{\text{policy}}(\theta^i_j)$ increases the probability of advantageous actions, 
\begin{align}\label{actor}
    \mathcal{L}_{\mathrm{policy}}(\theta^i_j)&=-\mathbb{E}_{B_j}\Bigg[\min \Bigg(\frac{\pi_{\theta^i_j}\left(a_k \mid s_k\right)}{\pi_{\hat{\theta}_j}\left(a_k \mid s_k\right)} A_k^{i}, \nonumber\\&\operatorname{clip}\Big(\frac{\pi_{\theta^i_j}\left(a_k \mid s_k\right)}{\pi_{\hat{\theta}_j}\left(a_k \mid s_k\right)}, 1-\epsilon, 1+\epsilon\Big) A_k^{i}\Bigg)\Bigg],
\end{align}
where $\epsilon$ is the clipping parameter, the clip function is used for stable training \cite{PPO}. $A_k^{i}$ is the estimated advantage function. Denote the discount factor as $\gamma$. Since intermediate inference rewards are all 0, $A_k^{i}$ can be expressed as
\begin{align}\label{advantage_function}
    A_k^{i}&=\gamma^{K-1-k}r^i_K-V_{\omega^i_j}(s_{K{-}1}).
\end{align}
$\mathcal{H}(\theta^i_j)$ is used to encourage action exploration,
\begin{align}\label{entropy}
\mathcal{H}\left(\theta_j^i\right)=\mathbb{E}_{s_k\sim \pi_{\hat{\theta}_j}}\left[-\sum_{a_k}\pi_{\theta_j^i}(a_k\mid s_k)\log{\pi_{\theta_j^i}(a_k\mid s_k)}\right].
\end{align}
$\mathcal{L}_{\text{value}}(\omega^i_j)$ is the loss for estimating the state value function,  
\begin{align}\label{critic}
\mathcal{L}_{\text{value}}(\omega^i_j)=\mathbb{E}_{B_j}\left[\left(V_{\omega^i_j}\left(s_{K-1}\right)-r^i_K\right)^2\right].
\end{align}

\subsubsection{Global Policy Optimization and Pareto Set Update}
The Kullback-Leibler (KL) divergence $\mathcal{D}_{\mathrm{KL}}(p\|q)$ quantifies the difference between the distribution $p$ and the distribution $q$, where $q$ is approximated to align with $p$. Hence, the global policy network $\hat{\theta}_{j}$ is updated to $\hat{\theta}_{j+1}$ by minimizing the following loss function,  
\begin{align}\label{global_policy}
\mathcal{L}_{\mathrm{KL}}(\hat{\theta}_{j+1})=&\sum_{i\in\{1,2\}}\mathcal{D}_{\mathrm{KL}}\left(\pi_{\theta_j^i}\left(\cdot \mid s_k\right) \| \pi_{\hat{\theta}_{j+1}}\left(\cdot \mid s_k\right)\right)\nonumber\\
&+\eta_3 \mathcal{D}_{\mathrm{KL}}\left(\pi_{\hat{\theta}_{j}}\left(\cdot \mid s_k\right) \| \pi_{\hat{\theta}_{j+1}}\left(\cdot \mid s_k\right)\right),
\end{align}
where $\eta_3$ is a positive hyperparameter. Consequently, the updated policy $\pi_{\hat{\theta}_{j+1}}$ is optimized to be close to both directional sub-policies $\pi_{\theta_j^i}$, while maintaining proximity to the policy from the previous iteration $\pi_{\hat{\theta}_{j}}$. 

 Based on policy $\pi_{\hat{\theta}_{j+1}}$, the Pareto solution set $\Lambda_j$ is updated as follows. Given the input $\boldsymbol{x}_T$ and $G_{\boldsymbol{w}}$, the policy $\pi_{\hat{\theta}_{j+1}}$ is used to infer a K-step solution $\hat{\boldsymbol{x}}_0^K$. 
If instance $G_{\boldsymbol{w}}$ has not yet appeared in the current Pareto solution set $\Lambda_j$, both $G_{\boldsymbol{w}}$ and $\hat{\boldsymbol{x}}_0^K$ are directly added to $\Lambda_j$. Otherwise, if the solution $\hat{\boldsymbol{x}}_0^K$ is identified as a non-dominated solution\cite{mo_ppo}, it is added to $\Lambda_j$. Simultaneously, any solutions in $\Lambda_j$ that are dominated by $\hat{\boldsymbol{x}}_0^K$ are removed from the set.

In general, the overall procedure of the proposed MO-CMPO algorithm is presented in Algorithm \ref{MO_CMPO_alg}.
\begin{algorithm}[t]
{\bf \small{Initialize:}} \small{consistency model network $\theta$, sub-policy networks $\theta^i$, value networks $\omega^i$, replay buffer $B_0=\varnothing$, Pareto solution set $\Lambda_0=\varnothing$}\\
{\bf \small{Parameters:}} \small{RL iteration number $J$, preference weight $\bm{w}$}
\caption{MO-CMPO}
\label{MO_CMPO_alg}
\begin{algorithmic}[1]
\STATE Supervised learning CM $\theta$ according to Algorithm \ref{CM_training_alg}, based on labeled Gurobi dataset
\STATE Initialize iteration $j=0$
\STATE Initialize the policy network $\hat{\theta}_0=\theta$

\REPEAT
    \REPEAT
        \STATE Sample problem $G$ from instances in Gurobi dataset,\\$\quad$sample preference weight $\boldsymbol{w}$: (\ref{weight_buffer}),\\$\quad$get embedded graph $G_{\boldsymbol{w}}$
        \STATE Perform $K$ step inference: (\ref{samplex_buffer}),\\$\quad$obtain feasible solution $\hat{x}_0^K$ by greedy algorithm,\\$\quad$decide $a_k$ based on $s_k$ according to $\pi_{\hat{\theta}_j}$,\\$\quad$get instantaneous rewards $\boldsymbol{r}_{k+1}$: (\ref{rewards})
        \STATE Store $(s_k, a_k, \boldsymbol{r}_{k+1})$ in the replay buffer $B_j$
    \UNTIL{$B_j$ is filled}
    \FOR{$i=1,2$}
        \STATE Update policy network $\theta^i_j$, value network $\omega^i_j$: (\ref{PPO})\text{-}(\ref{critic})
    \ENDFOR
    \STATE Update global policy $\hat{\theta}_j$ to $\hat{\theta}_{j+1}$: (\ref{global_policy})
    \STATE Update Pareto solution set $\Lambda_j$ to $\Lambda_{j+1}$ based on policy $\pi_{\hat{\theta}_{j+1}}$
    \STATE Update iteration $j=j+1$
\UNTIL{$j=J$}
\end{algorithmic}
{\bf \small{Output:}} policy network $\hat{\theta}_{J}$, Pareto solution set $\Lambda_J$
\end{algorithm}
 
\section{Simulation Results}\label{simulation_results}
\subsection{Environment Settings}
We consider a system consisting of $|\mathcal{M}|=20$ MEC servers that collectively support up to $|\mathcal{U}|=50$ VR users. The number of available virtual spaces is set to $|\mathcal{V}|=10$. For each virtual space, one parameter value is randomly selected from predefined level sets. Specifically, the cache size levels are given by $c_v\in\{10, 50, 100, 500, 1000, 1500\}$ $\mathrm{MB}$, the basic energy consumption levels by $E_v \in [10, 15, 20, 30, 40, 50]$ $\mathrm{J}$, the FLOP levels by $h_v \in\{20, 50, 80, 100, 120, 150\}$ $\mathrm{M\ cycles}$, and the data size levels by $D_v \in\{10, 25, 50, 100, 120, 150\}$ $\mathrm{Mbits}$. The probability $p_{u,v}$ at each time slot is derived using the positional VR track dataset in \cite{VRtrack}, following the methodology in \cite{VRgame_Zhu}. For each MEC server $m$, the cache capacity $C_m$ is uniformly distributed between 15 GB and 20 GB, the computation frequency $F_m$ ranges from 2 $\mathrm{GHz}$ to 5 $\mathrm{GHz}$, and the maximum task limit $N_m$ is randomly assigned within the interval $[10,15]$. 

The NR BS data in MAN are obtained from the OpenCellID database. Two datasets, corresponding California to Virginia, provide geographic coordinates and user counts for individual cells. In California, 500 cells with more than 50 users are selected, with inter-cell distances ranging from 18 m to 65.2 km (median: 7 km). In Virginia, 1,600 cells with more than 50 users are selected, with distances ranging from 14 m to 168.6 km (median: 48.4 km). For each dataset, 20 cells are randomly sampled to serve as MEC server locations. The energy coefficient is set to $10^{-25}$ $\mathrm{s}\cdot \mathrm{J/cycle}$. The latency and energy consumption associated with synchronization and sensor information are assumed to scale linearly with the inter-cell distance, with proportionality constants of $0.1$ $\mathrm{ms/km}$ and $0.15$ $\mathrm{J/km}$, respectively, thereby generating the parameters $d_{m,n}, e_{m,n}, d_{u,m}, e_{u,m}$. For viewport frames, latency and energy consumption are assumed to be proportional to both the distance and the data size, with proportionality constants of $0.06$ $\mathrm{ms/(km\cdot Mbit)}$ and $0.01$ $\mathrm{J/(km\cdot Mbit)}$, respectively, which define the parameters $\kappa_{m,u}, \zeta_{m,u}, \kappa_{u}, \zeta_{u}$.

\subsection{MO-CMPO Training}
For supervised learning with CM, the number of diffusion steps is set to $T=1000$ with $\alpha=0.2$. The Adam optimizer is employed with an initial learning rate of $1\times10^{-4}$, following a cosine decay schedule with a base rate of $1\times10^{-4}$. L2 regularization is applied with a weight decay coefficient of $2\times10^{-4}$. The sparse GNN consists of 12 layers with a hidden dimension of 256, where information aggregation is performed via simple summation. By solving problem $\mathcal{P}1$ using Gurobi, with total latency and total energy as the single optimization objective, we generate 81,920 labeled data samples from the California and Virginia datasets.

For fine-tuning with RL, the number of inference steps is set to $K=3$. The entropy loss weight is $\eta_1=0.01$, and the critic network mean square error (MSE) loss weight is $\eta_2=0.5$. The clipping parameter  is set to $\epsilon=0.2$, and the discount factor in the advantage estimation function is $\gamma=0.99$. The replay buffer size is $|B_j|=50,000$, and the number of iterations is  $J=50$. All experiments are conducted on eight NVIDIA GeForce RTX 4090 GPUs. An environment, referred to as SCCEnv, is developed for agent training, built upon the Gym framework\cite{gym}.

\subsection{Performance Evaluation}
\begin{figure}[t]
\centering{\includegraphics[width=0.45\textwidth]{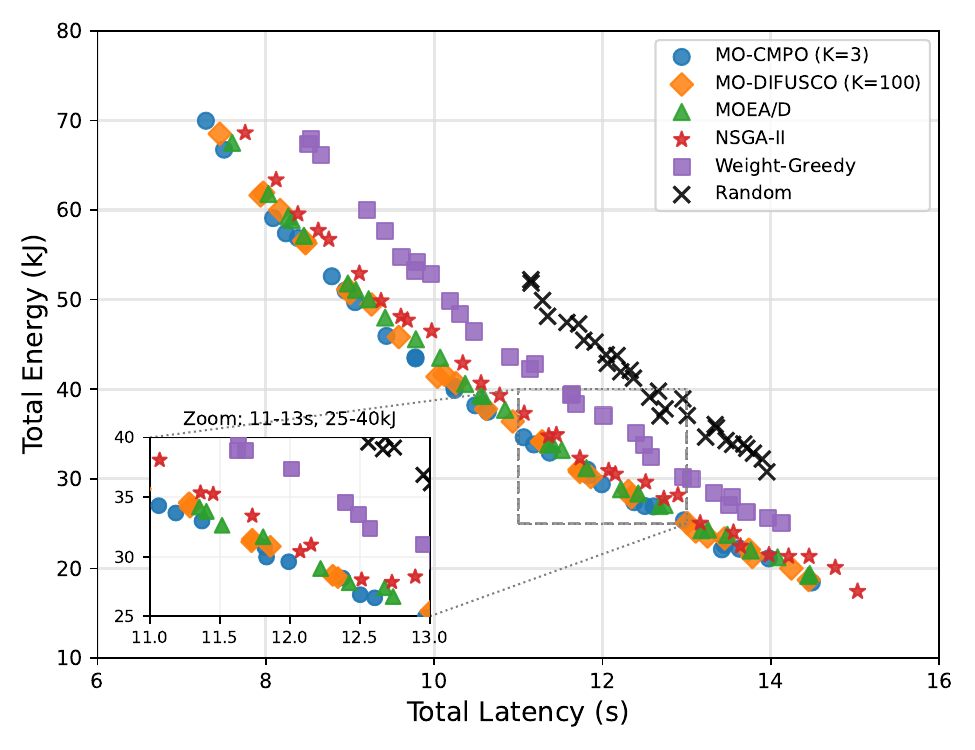}}
\caption{Pareto fronts with different algorithms using an instance, where average inter-cell distance is 6.856 $\mathrm{km}$, average $c_v$ is 608 $\mathrm{MB}$, average $E_v$ is 29 $\mathrm{J}$, average $h_v$ is 88 $\mathrm{M}$ $\mathrm{cycles}$, average $D_v$ is 56 $\mathrm{Mbit}$, from California dataset.}
\label{pareto_ca}
\end{figure}

\begin{figure}[t]
\centering{\includegraphics[width=0.45\textwidth]{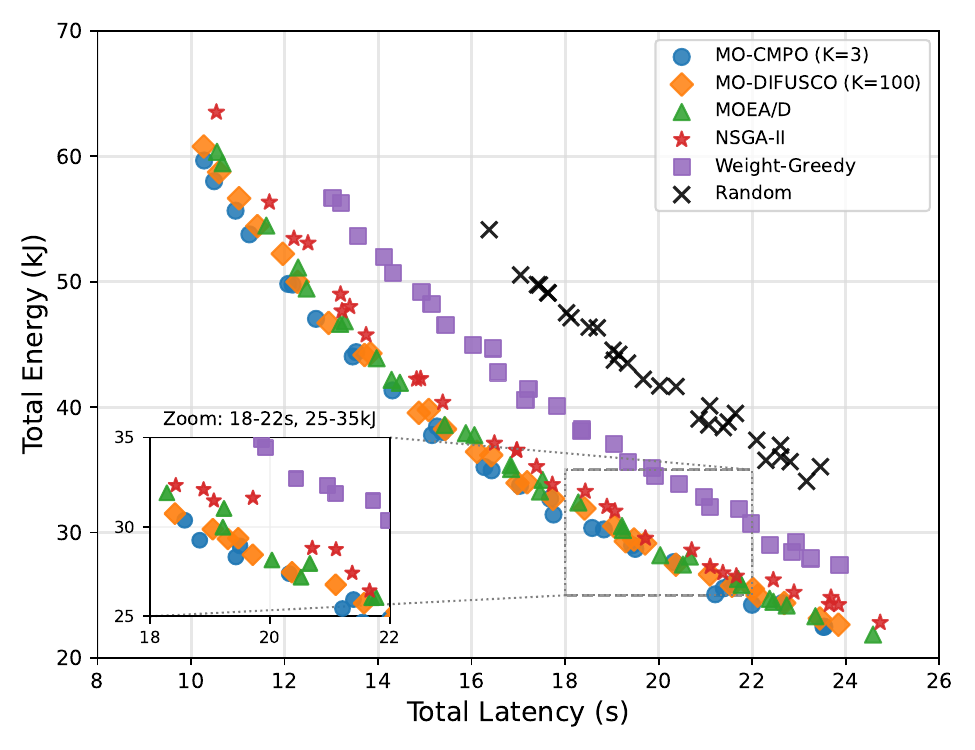}}
\caption{Pareto fronts with different algorithms using an instance, where average inter-cell distance is 43.833 $\mathrm{km}$, average $c_v$ is 283 $\mathrm{MB}$, average $E_v$ is 26 $\mathrm{J}$, average $h_v$ is 83 $\mathrm{M}$ $\mathrm{cycles}$, average $D_v$ is 81 $\mathrm{Mbit}$, from Virginia dataset.}
\label{pareto_va}
\end{figure}

 The Pareto frontiers obtained by different algorithms are illustrated in Fig. \ref{pareto_ca} and Fig. \ref{pareto_va}. Five baseline algorithms are considered for comparison with the proposed MO-CMPO: (i) \textbf{MO-DIFUSCO ($K=100$)}, which follows the same training pipeline as MO-CMPO, i.e., incorporating preference weights with a two-stage procedure of supervised learning followed by RL fine-tuning. The difference lies in that MO-DIFUSCO (K=100) does not employ the CM algorithm but instead adopts DIFUSCO\cite{DIFUSCO}, a diffusion model with 100 inference steps. (ii) \textbf{MOEA/D}, a representative multi-objective evolutionary algorithm \cite{MOEA/D}. (iii) \textbf{NSGA-II}, a widely used multi-objective genetic algorithm \cite{NSGA-II}. (iv) \textbf{Weight-Greedy}, which converts the multi-objective problem into a single-objective problem via $T(\cdot) + \ell E(\cdot)$, followed by a greedy optimization, where $\ell \in \{0.1, 0.5, 1, 2, 4, 8, 10\}$. (v) \textbf{Random}, which generates feasible strategies uniformly at random. Fig. \ref{pareto_ca} and Fig. \ref{pareto_va} correspond to instances from the California and Virginia datasets, respectively. To reflect representative scenarios, the selected instances exhibit average inter-cell distances close to the median values of their respective datasets. Both MO-DIFUSCO (K=100) and MO-CMPO (K=3) employ DMs and their variants to approximate the distribution of optimal solutions. The stochasticity inherent in the generated trajectories introduces exploration during RL fine-tuning, thereby contributing to their superior performance.

\begin{figure}
\centering{\includegraphics[width=0.45\textwidth]{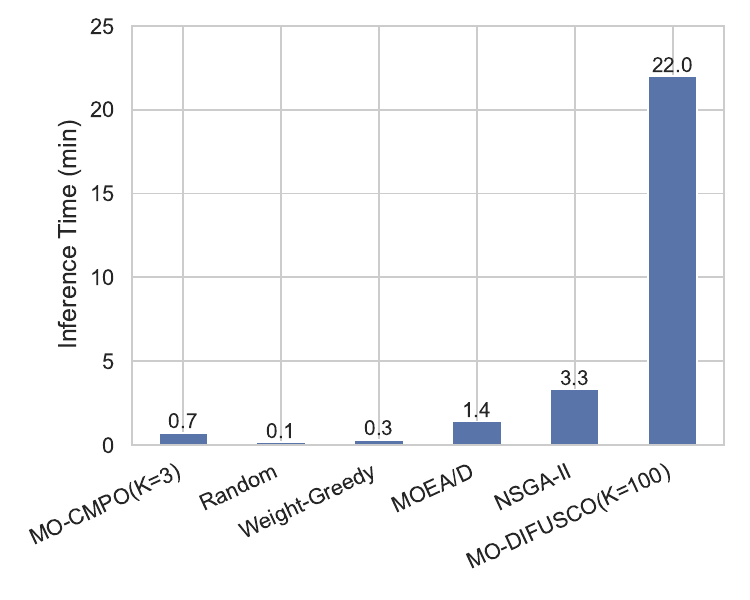}}
\caption{Inference time of different algorithms. }
\label{time}
\end{figure}

The average inference latency per instance for different algorithms is shown in Fig. \ref{time}. It can be observed that the inference latency of MO-CMPO (K=3) remains on the order of seconds, comparable to that of the low-complexity Random and Weight-Greedy algorithms. In contrast, the MO-DIFUSCO (K=100) algorithm incurs significantly higher inference latency, primarily due to the large number of inference steps, which is consistent with the behavior of conventional diffusion models. Owing to the use of the CM algorithm, MO-CMPO achieves comparable or superior performance with substantially fewer inference steps, thereby reducing inference time while maintaining solution quality.

To further validate the performance advantage of the proposed algorithm, the results are summarized in Table \ref{performance}. The columns HV-CA and HV-VA denote the normalized hypervolume (HV) obtained from the California and Virginia datasets, respectively, while HV represents the average normalized hypervolume across all instances. The last column report the inference latency of each algorithm. The normalized hypervolume is defined as follows. Since HV measures the portion of the objective space dominated by the solution set, a reference point $[T_{ref},E_{ref}]$ must first be specified. This point is guaranteed to be dominated by all feasible solutions and is set to $[50, 100]$ in this paper. For solutions with total latency $\tilde{T}(\cdot)$ and total energy $\tilde{E}(\cdot)$, normalization is performed with respect to be the reference point, i.e., $\tilde{T}(\cdot) = T(\cdot)/T_{ref}$ and $\tilde{E}(\cdot) = E(\cdot)/E_{ref}$. The normalized HV is then formulated as
\begin{align}\label{HV}
\operatorname{HV}_{norm}(\mathcal{S}_{G}, \mathbf{1})=\operatorname{Volume}\left(\bigcup_{[\tilde{T}, \tilde{E}] \in \mathcal{S}_{G}}\left[\tilde{T}, 1\right] \times\left[\tilde{E}, 1\right]\right),
\end{align}
 where $\mathcal{S}_G$ denotes the Pareto solution set of the instance $G$. As summarized in Table \ref{performance}, MO-CMPO(K=3) achieves the largest hypervolume, thereby demonstrating the best performance, while the Random algorithm yields the smallest HV, consistent with the observations in Fig. \ref{pareto_ca} and Fig. \ref{pareto_va}. A subtle difference is that, in Fig. \ref{pareto_ca} and Fig. \ref{pareto_va}, MO-CMPO(K=3) is closest to the lower-left corner, followed by MO-DIFUSCO(K=100), MOEA/D, and NSGA-II. In contrast, the HV-CA and HV-VA values in Table \ref{performance} for MO-DIFUSCO(K=100), MOEA/D, and NSGA-II do not strictly follow this order. This discrepancy can be attributed to the fact that HV captures not only the proximity of the solution set to the optimal front,  but also its coverage, which may induce variations in the HV metric across instances.

As shown in table and figures, the proposed MO-CMPO consistenly outperforms the baseline methods, yielding a Pareto front with a more uniform distribution. Moreover, MO-CMPO attains relatively low inference latency while maintaining strong solution quality. It is further observed that, for the same algorithm, the HV-VA values in Table \ref{performance} are generally larger than the corresponding HV-CA values. A similar trend is evident in the scatter points, where the Pareto fronts in Fig. \ref{pareto_va} exhibit broader coverage compared to those in Fig. \ref{pareto_ca}. This phenomenon can be attributed to the larger average inter-cell distance in the Virginia dataset, which results in a wider range of latency and energy values and, consequently, a larger optimization space between the two objectives.

\begin{table}
\centering
\caption{{Performance of different algorithms.}}
\label{performance}
\begin{tabular}{l!{\vrule}cccc}
\toprule
Algorithm & HV-CA & HV-VA & HV & time \\ \midrule
Random   & 0.5323 & 0.4304 & 0.4823  & \textbf{9s} \\
Weight-Greedy  & 0.6008 & 0.5096 & 0.5559  & 17s \\
NSGA-II    & 0.6615 & 0.5653 & 0.6163  & 3min18s \\
MOEA/D & 0.6572 & 0.5744 &  0.6165 & 1min24s \\ 
MO-DIFUSCO(K=100) & 0.6622 & 0.5742  & 0.6187  & 22min \\
MO-CMPO(K=3)    & \textbf{0.6659} & \textbf{0.5771} &  \textbf{0.6221} & 43s \\ \bottomrule
\end{tabular}
\end{table}

\subsection{Parameter Influence}
\begin{figure*}[t]
    \centering
    \begin{subfigure}{0.45\textwidth}
        \centering
        \includegraphics[width=\linewidth]{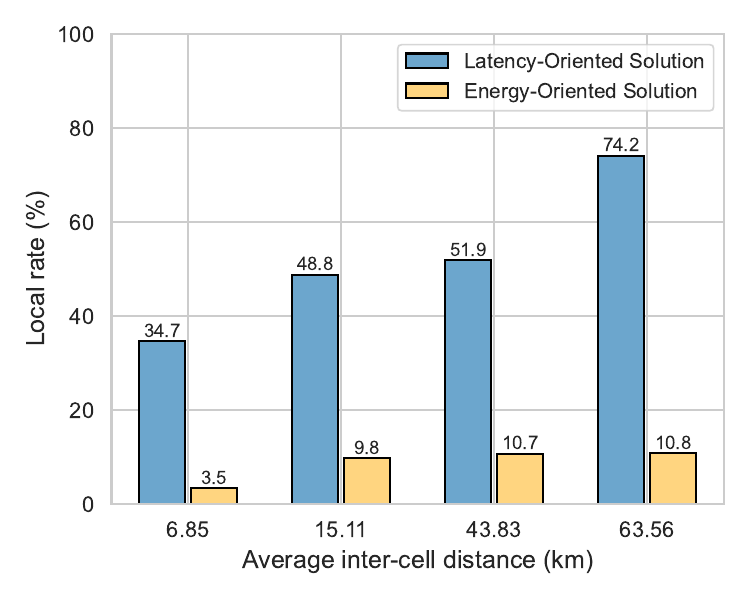}
        \caption{Local rate for different average inter-cell distances and orients.}
        \label{local_rate}
    \end{subfigure}
    \hfill
    \begin{subfigure}{0.45\textwidth}
        \centering
        \includegraphics[width=\linewidth]{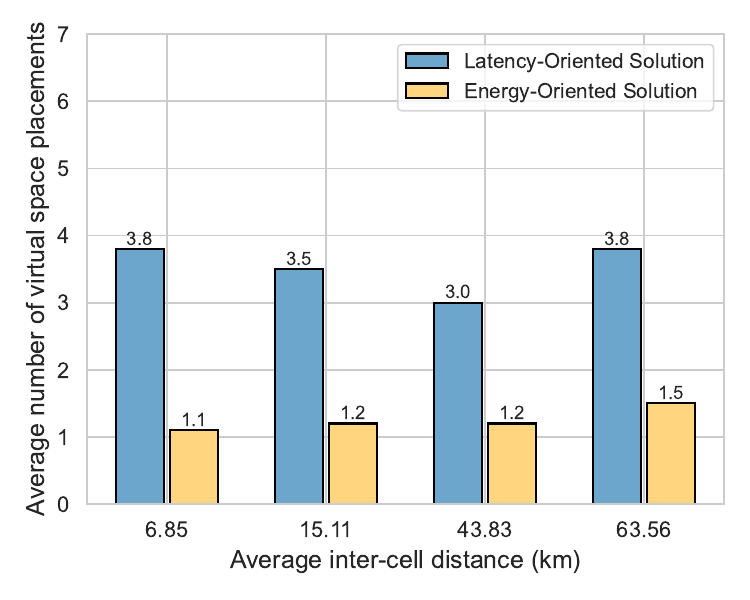}
        \caption{Average number of virtual space placements for different average inter-cell distances and orients.}
        \label{placement}
    \end{subfigure}
    
    \vspace{0.5em} 

    \begin{subfigure}{0.45\textwidth}
        \centering
        \includegraphics[width=\linewidth]{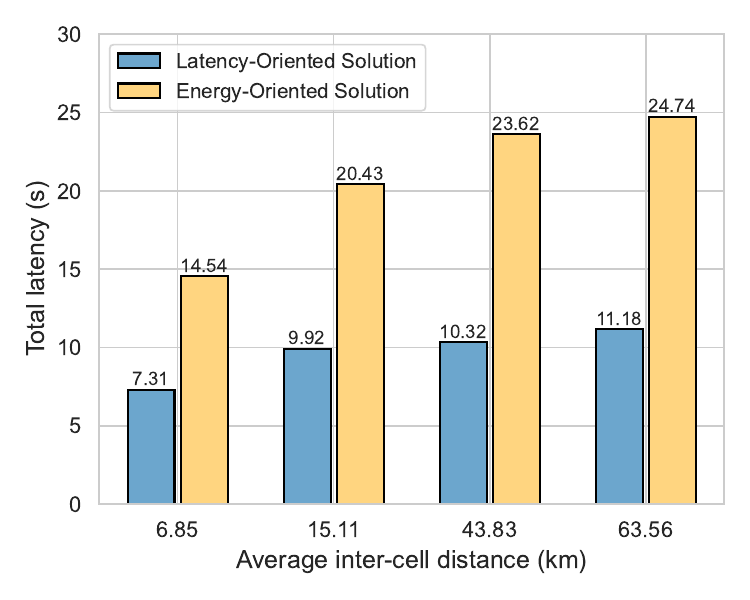}
        \caption{Total latency for different average inter-cell distances and orients.}
        \label{latency}
    \end{subfigure}
    \hfill
    \begin{subfigure}{0.45\textwidth}
        \centering
        \includegraphics[width=\linewidth]{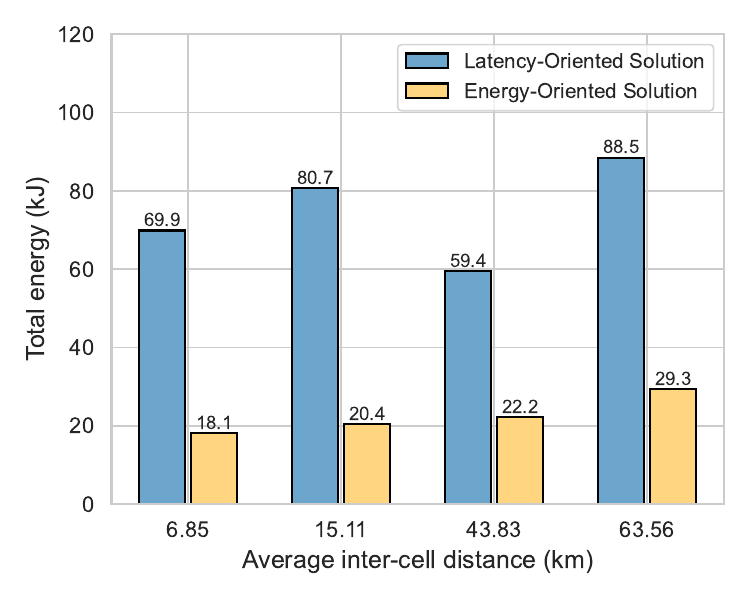}
        \caption{Total energy for different average inter-cell distances and orients.}
        \label{energy}
    \end{subfigure}

    \caption{Effect of parameter variation on strategy.}
    \label{parameter influence}
\end{figure*}

Based on the proposed MO-CMPO, we further investigate the relationship between policy and  system parameters. We define a \textit{latency-oriented solution} and an \textit{energy-oriented solution} as $\boldsymbol{x}=\hat{\boldsymbol{x}}_0^K$ obtained by the MO-CMPO algorithm when the preference weight is set to $\boldsymbol{w}=[1,0]$ and $\boldsymbol{w}=[0,1]$, respectively. The impact of parameter variations on the learned policy is illustrated in Fig. \ref{parameter influence}, where instances are arranged from left to right with progressively increasing average inter-cell distances, while other parameters remain fixed. 

In Fig. \ref{local_rate}, we evaluate the local rate, which measures the proportion of users selecting their local MEC among all user–virtual space pairs, thereby reflecting the inclination of users to execute tasks locally. Let the local MEC index of user $u$ be given by $\mathbf{lp}=\left(l p_1, \ldots, l p_U\right) \in \mathcal{M}^{\mathcal{U}}$. The MEC selection of user 
$u$ in virtual space $v$ is represented by $x_{u,v,m}=1$ and $\sum_{m \in \mathcal{M}} x_{u, v, m} \in\{0,1\}$. If user $u$ in virtual space $v$ selects its local MEC, the corresponding local state is $x_{u, v, lp_u}, \forall u \in \mathcal{U}, v \in \mathcal{V}$. The local rate is thus defined as the proportion of localized selections among all user-virtual space selections,
\begin{align}\label{local_rate_eq}
    \frac{\sum_{u \in \mathcal{U}} \sum_{v \in \mathcal{V}} x_{u, v, lp_u}}{\sum_{u \in \mathcal{U}} \sum_{v \in \mathcal{V}} \sum_{m \in \mathcal{M}} x_{u, v, m}}.
\end{align}
As shown, the local rate increases with the average inter-cell distance for both latency-oriented and energy-oriented solutions. Furthermore, the latency-oriented solution consistently achieves a significantly higher local rate than the energy-oriented solution, indicating that assigning tasks to the user's local MEC is particularly effective for reducing overall latency.

In Fig. \ref{placement}, the average number of virtual space placements represents the number of MECs that simultaneously host the same virtual space selected by users. This metric captures scenarios where multiple distributed MECs execute the same task in parallel, and is denoted as,
\begin{align}\label{placement_eq}
     \frac{1}{|\mathcal{V}|} \sum_{v \in \mathcal{V}}\left|\left\{m \in \mathcal{M} \mid \sum_{u \in \mathcal{U}} x_{u, v, m}>0\right\}\right|,
\end{align}
where the outer cardinality operator $|\cdot|$ denotes the number of distinct MECs. It can be observed from Fig. \ref{placement} that for latency-oriented tasks, the average number of virtual space placements does not exhibit a clear trend with increasing inter-cell distance. For energy-oriented tasks, a slight upward tendency is observed, though the effect is weak. Furthermore, the average number of virtual space placements is consistently higher for the latency-oriented solution than for the energy-oriented ones, which aligns with the observations in Fig. \ref{local_rate}. Specifically, because latency-oriented users tend to execute tasks on their local MECs, the same virtual space must often be placed on multiple MECs, thereby increasing the number of placements.

The impact of these placements on total latency and total energy is shown in Fig. \ref{latency} and Fig. \ref{energy}. As illustrated, latency-oriented tasks achieve substantially lower total latency compared to energy-oriented tasks, whereas the opposite holds for total energy. In Fig. \ref{latency}, both latency- and energy-oriented solutions exhibit higher total latency as the average inter-cell distance increases. In contrast, Fig. \ref{energy} shows that the total energy consumption of latency-oriented solutions has no consistent correlation with inter-cell distance. This behavior arises because energy consumption is influenced not only by user–MEC distance but also by the number of virtual space placements. For example, when the average inter-cell distance is $43.83$ km, Fig. \ref{placement} shows relatively few placements for latency-oriented tasks, which corresponds to the reduced total energy observed at the same point in Fig. \ref{energy}.

\section{Conclusion}\label{conclusion}
In this paper, we proposed an SCC framework to enable multi-user VR services over distributed MEC networks. By jointly modeling physical and virtual spaces, we formulated a MOCO problem that captures both latency and energy consumption. To tackle the inherent complexity, we developed MO-CMPO, a novel algorithm that combines the fast inference capability of consistency models with reinforcement learning-based fine-tuning. Extensive simulations on real-world BS datasets that MO-CMPO outperforms conventional heuristics, evolutionary algorithms, and diffusion-based methods, achieving superior Pareto front quality with substantially reduced inference latency. Moreover, our analysis revealed the trade-offs between local MEC deployment, synchronization costs, and cross-MEC transmissions, offering new insights into resource allocation for immersive applications. Future research will focus on extending the SCC framework to support a wider range of immersive media services and heterogeneous MEC environments.

\bibliographystyle{IEEEtran}
\bibliography{reference}

\end{document}